\documentclass[aps,showpacs,twocolumn]{revtex4}
\usepackage{amsfonts}
\usepackage{amsmath}
\usepackage{amssymb}
\usepackage{graphicx}
\usepackage{bm}
\usepackage{color}

\input{tcilatex}
\begin{document}

\title{Specific Salt Effects on Thermophoresis of Charged Colloids}
\author{Kyriakos A. Eslahian$^{1,2}$, Arghya Majee$^{3,4,5}$, Michael Maskos$%
^{2}$, Alois W\"{u}rger$^{5}$}
\affiliation{$^{1}$Bundesanstalt f\"{u}r Materialforschung und Pr\"{u}fung, Unter den
Eichen 87, 12205 Berlin, Germany}
\affiliation{$^{2}$Institut f\"{u}r Mikrotechnik Mainz GmbH, Carl-Zeiss-Strasse 18-20,
55129 Mainz, Germany}
\affiliation{$^{3}$Max-Planck-Institut f\"{u}r Intelligente Systeme, Heisenbergstrasse 3,
70569 Stuttgart, Germany}
\affiliation{$^{4}$Institut f\"{u}r Theoretische Physik IV, Universit\"{a}t Stuttgart,
Pfaffenwaldring 57, 70569 Stuttgart, Germany}
\affiliation{$^{5}$Laboratoire Ondes et Mati\`{e}re d'Aquitaine, Universit\'{e} de
Bordeaux \& CNRS, 351 cours de la Lib\'{e}ration 33405 Talence, France}

\begin{abstract}
We study the Soret effect of charged polystyrene particles as a function of
temperature and electrolyte composition. As a main result we find that the
Soret coefficient is determined by charge effects, and that non-ionic
contributions are small. In view of the well-kown electric-double layer
interactions, our thermal field-flow fractionation data lead us to the
conclusion that the Soret effect originates to a large extent from
diffusiophoresis in the salt gradient and from the electrolyte Seebeck
effect, both of which show strong specific-ion effects. Moreover, we find
that thermophoresis of polystyrene beads is fundamentally different from
proteins and aqueous polymer solutions, which show a strong non-ionic
contribution.
\end{abstract}

\maketitle

\section{Introduction\textit{\ }}

When applying a temperature gradient to a colloidal suspension, molecules
and nanoparticles migrate to the cold or to the hot, depending on the solute
and solvent properties \cite{Wie04,Pla06,Pia08,Wue10}. In recent years, this
Soret effect, or thermophoresis, has been used for colloidal confinement to
a micron-sized hot spot in a channel or thin film \cite{Duh06,Jia09,Mae11},
and for self-propulsion of hot Janus particles in active colloids \cite%
{Jia10,But12, Qia13}. The underlying thermal forces are very sensitive to
the electrolyte composition. Thus colloids move to the cold side in NaCl
solution and to the hot in NaOH, because of the Seebeck effect which takes
opposite signs in these electrolytes \cite{Put05,Wue08,Vig10,Maj11}. The
related electric field has been discussed in view of translocating DNA
through nanopores \cite{He13} and accumulating charge in a micron-size hot
spot \cite{Maj12,Maj13}. More generally, thermophoresis shows specific-ion
effects \cite{Esl12,Lan12} similar to the Hofmeister series of protein
interactions. As another signature for charge effects, the electrostatic
repulsion of nearby particles gives rise to a characteristic dependence on
volume-fraction and salinity \cite{Pia02,Fay05}

The temperature dependence of thermophoresis is poorly understood, in
particular the strong increase with $T$ and the change of sign that occur in
many instances \cite{Iac03,Iac06,Put07,Won12,Koe13}.\ In a study on lysozyme
protein, Iacopini \textit{et al.} separated the effect of temperature and
salinity \cite{Iac03}; their data give clear evidence for an important
non-ionic interaction that dominates the temperature dependence. This has
been confirmed for non-ionic surfactants \cite{Iac06,Nin08} and PNIPAM
microgels \cite{Won12,Koe13}. A phenomenological law proposed by Piazza
describes remarkably well the behavior of non-ionic solutes, proteins,
polyelectrolytes, and charged latex beads \cite{Iac06}. Regarding the
latter, there is so far no direct proof for a non-ionic driving mechanism.
To know whether it exists or not, is important for an efficient surface
functionalization of self-propelling colloids.

\begin{figure}
\includegraphics[width=\columnwidth]{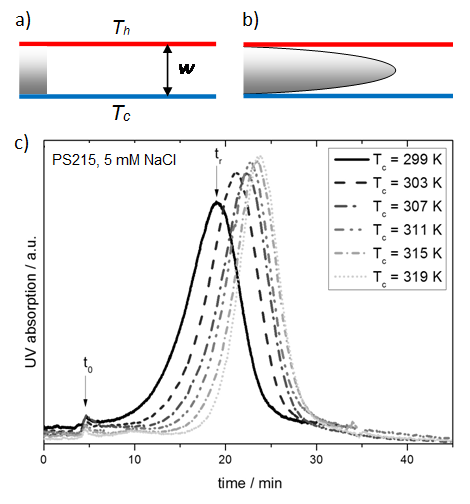}
\caption{Schematic view of our ThFFF
setup. a) The Soret equilibrium state in the vertical temperature difference
is prepared at the entrance of the channel, with a vertical concentration
profile $c=c_{0}e^{-z/\ell }$. b) Pumping the suspension through the channel
results in a parabolic velocity profile. At the exit, the colloidal content
is determined by detection of the UV-absorption as a function of time. c)
Typical elution profiles consist of a strong Gaussian feature at $t=t_{r}$
that describes the arrival of the particles accumulated close to the lower
wall.\ We show elution patterns at different temperatures. At higher $T$,
the position of the Gaussian is shifted to later times, implying a smaller
retention length $\ell $ and thus a higher Soret coefficient. }
\label{fig1}
\end{figure}

The present paper addresses two questions on thermophoresis of charged
particles: First, is there a non-ionic contribution and, second, what is the
origin of the temperature dependence? We report on thermal field-flow
fractionation (ThFFF) measurements of the Soret coefficient of polystyrene
sulfonate (PS) beads as a function of electrolyte composition and
temperature. In view of the theory of phoretic motion of colloids, our data
strongly suggest that the answer to the above questions is intimately
related to the temperature-induced salt-ion gradients.

The paper is organized in the following way. Sect.\ 2 gives details of our
experimental setup, and Sect. 3 an overview of the theoretical description
of thermophoresis due to electric-double layer forces. The effects of the
electrolyte composition and temperature are discussed in Sects.\ 4 and 5. We
conclude in Sect.\ 6 with a discussion of our main results and a comparison
to protein thermophoresis.

\section{Experimental section}

We briefly present the principle of ThFFF as shown in Fig.\ 1.\ The
colloidal suspension is injected at the channel entrance and, after an
equilibration time of one minute \cite{Mic02}, reaches a steady state with
vertical concentration profile $c=c_{0}e^{-z/\ell }$, where the retention
length $1/\ell =S_{T}\Delta T/w$ depends on the Soret coefficient $S_{T}$,
the temperature difference $\Delta T=T_{h}-T_{c}$ across the channel, and
its width $w$. In our setup the channel length is 46.4 cm, the breadth 1.8
cm, and the width $w=$100 $\mu $m.\ Pumping the suspension through the
channel results in a parabolic velocity profile. At the exit, the colloidal
content is determined by detection of the UV-absorption as a function of
time. Typical elution profiles as in Fig. 1c, consist of a strong Gaussian
feature at $t=t_{r}$ that describes the arrival of the particles accumulated
close to the upper or lower wall; the small precursor peak occurs at the
moment where the front of the parabolic flow profile reaches the exit.\ At
higher $T$, the position of the Gaussian is shifted to later times, implying
a smaller retention length $\ell $ and thus a higher Soret coefficient. In
all experiments the flow rate is 0.3 mL/min and the colloidal mass fraction
at the moment of injection is in the order of 10$^{-4}$; the temperature
difference is in the range $\Delta T=15...25$ K for the data shown in Fig.\
2 and 3, and $\Delta T=10...15$ K for those in Figs.\ 4 and 5. The effective
temperature is calculated by assuming a linear temperature profile through
the channel and taking the concentration-weighted mean value. For positive
and negative $S_{T}$ this gives $T=T_{c}+\Delta T(\ell /w)$, and $%
T=T_{h}-\Delta T(\ell /w)$, respectively.

The Soret coefficient is determined following a standard procedure \cite%
{Mic02}. The channel void time $t_{0}$, which is given by the centre of
gravity of the elugram of one non-retained sample, is divided by the
retention time to calculate the retention ratio $R=t_{0}/t_{r}$. The almost
parabolic velocity profile results in a characteristic elution pattern that
allows us to determine the length $\ell $ through the relation $R=6\lambda
\coth (1/2\lambda )-12\lambda ^{2}$, where $\lambda =\ell /w=1/S_{T}\Delta T$
\cite{Sch00}. For most of our data, $\lambda $ is in the range from 2 to
6\%. In the calculated retention parameters, we account for deviations of
the flow profile due to the temperature dependent viscosity. Retention
parameters have to be corrected in terms of the deviation parameter \cite%
{Hel94}, taking the not ideal flow profile into account, which results from
the temperature dependence of the solvent \cite{Gun84}. ThFFF does not
distinguish whether the colloid accumulates at the upper or lower plate, and
thus provides only the absolute value $|S_{T}|$ of Soret coefficient. Thus
the sign has to be inferred from continuity of the data and from previous
experiments.

The samples PS 90 and PS 136 are prepared by surfactant-free emulsion
polymerization, according to Juang and Krieger \cite{Jua76}. Two-phase
polymerization is performed in a glass reactor, thus ensuring optimized
stirring and temperature control. After heating the reactor up to 65%
${{}^\circ}$%
C, 177.7 g of water and 56.8 g of styrene are added. The difference between
preparation of PS 90 and PS 136 is the amount of ionic stabilizing comonomer
introduced: 0.63 g (PS 90) and 0.41 g (PS 136) of 4-styrene sulfonate. After
5 min of homogenization, the radical initiator potassium persulfate, 0.36 g
solved in 10 g of water, is added drop-wise. The reaction is kept alive for
24 h, and then latex emulsions are cooled down to room temperature, filtered
and dialyzed. The stability of resulting colloidal suspensions is given by
electrostatic repulsion, caused exclusively by the introduced sulfonate
groups. The sample PS 215 is purchased from Thermo Fisher Scientific
(Waltham, MA, USA), and used without further treatment.

All experiments are performed in de-ionized water ($18.2\times 10^{4}\Omega $%
m) without adding surfactant. Except styrene which has to be distilled in
prior to the synthesis, all chemicals are introduced as purchased in
analytical grade from Merck (Darmstadt, Germany). Hydrodynamic radii of the
particles used in this report are determined by dynamic multi-angle light
scattering and the spherical shape is verified by transmission electron
microscopy. All samples reveal having a very low polydispersity index.

\section{Charge-driven thermophoresis}

Charge effects on colloidal motion in a non-uniform electrolyte solution
arise from the interaction with the non-uniform electric double-layer \cite%
{Fay08}. In our systems the Debye screening length $\kappa ^{-1}$ is small
as compared to the particle size, and the surface potential $\zeta $ takes
moderate values. Thus viscous effects can be treated in boundary layer
approximation, and the screened electrostatics can be evaluated to second
order in $\zeta $. Then the particle velocity reads \cite{Wue10} 
\begin{equation}
u=-\frac{\varepsilon \zeta ^{2}}{12\eta }\left( \frac{\nabla T}{T}-\frac{%
\nabla n_{0}}{n_{0}}-\frac{\nabla \varepsilon }{\varepsilon }\right) +\frac{%
\varepsilon \zeta }{\eta }E,  \label{2}
\end{equation}%
with the solvent viscosity $\eta $, permittivity $\varepsilon $, and
salinity $n_{0}$, and an applied electric field $E$. The terms in
parentheses describe motion in a temperature gradient \cite{Ruc81}, in a
salt gradient \cite{And89}, and due to an electrostrictive force \cite{Hua08}%
; the particle moves toward lower temperature but higher salinity and
permittivity. The last term in (\ref{2}) accounts for electrophoresis in an
electric field $E$, with the Helmholtz-Smoluchowksi mobility $\varepsilon
\zeta /\eta $.

Here we deal with an applied thermal gradient and its companion fields
induced by the temperature dependent properties of the electrolyte solution
according to%
\begin{equation}
\tau =-\frac{T}{\varepsilon }\frac{d\varepsilon }{dT},\ \ \ \alpha =-\frac{T%
}{n_{0}}\frac{dn_{0}}{dT},\ \ E=S\nabla T.  \label{4}
\end{equation}%

Parameter $\tau $ accounts for the temperature-dependent permittivity of
water. The solvation enthalpies of positive and negative salt ions vary with
temperature; they result in a non-uniform salinity with thermal diffusion
factor $\alpha $ and a thermoelectric field with Seebeck coefficient $S$.
With the quantities defined in (\ref{4}) the particle velocity reads $%
u=-D_{T}\nabla T$, where\ the companion fields are absorbed in the mobility $%
D_{T}$. Its ratio with the Stokes-Einstein coefficient $D=k_{B}T/6\pi \eta a$
gives the Soret coefficient $S_{T}=D_{T}/D$, which varies linearly with the
particle radius $a$. From the above equations we have 
\begin{equation}
S_{T}=\frac{6\pi a}{k_{B}T^{2}}\left( \frac{\varepsilon \zeta ^{2}}{12}%
(1+\tau +\alpha )-\varepsilon \zeta ST\right) ,  \label{6}
\end{equation}%
where the expressions in parentheses have the dimension of a force, and take
typical values of a few pN. A positive sign of $S_{T}$ means that the
particle moves to lower temperatures.

It is instructive to compare the magnitude of the different contributions to 
$S_{T}$. With $a=215$ nm in the overall prefactor and $\zeta =-50$ mV, we
find that Ruckenstein's thermo-osmotic term, that is, the \textquotedblleft
1\textquotedblright\ in Eq. (\ref{6}) contributes 0.1 K$^{-1}$, and the
electrostrictive force $\propto \tau $ about 0.2 K$^{-1}$. In NaCl solution,
the specific-ion effects with coefficients $\alpha $\ and $S$ contribute 0.4
and 0.6 K$^{-1}$, respectively; in other words, the four terms are positive
and of the same order of magnitude. In NaOH solution, however, the
ion-specific effects contribute 0.6 and $-3.5$ K$^{-1}$. These numbers imply
that the Soret effect is to a large extent determined by the Seebeck and
diffusiophoresis contributions. For negatively charged particles in NaOH
solution, one expects the Seebeck term to result in a negative Soret
coefficient \cite{Wue08}.

Colloidal particles dispersed in an electrolyte solution constitute a
quaternary system. An effective binary system as in Eq. (\ref{6}) is
obtained by eliminating the salt ion concentrations $n_{\pm }$ as variables;
thus\ the $\zeta $-potential accounts for well-known composition of the
electric-double layer. The parameters $\alpha $ and $S$\ depend on the
non-uniform salinity and the surface charges related to the Seebeck
thermopotential. This reduction is achieved by solving the diffusion
equation for thermally driven ion currents and by replacing $\nabla n_{\pm }$
with their stationary values \cite{Mor99,Wue10}. This relies in particular
on the fact that the salt relaxation times are much shorter than that of the
colloidal distribution function. Moreover, at sufficiently small
colloidal volume fractions, the back-reaction of the colloidal macro-ions 
on $\alpha $ and $S$ can be neglected \cite{Maj11}.

\begin{figure}
\includegraphics[width=\columnwidth]{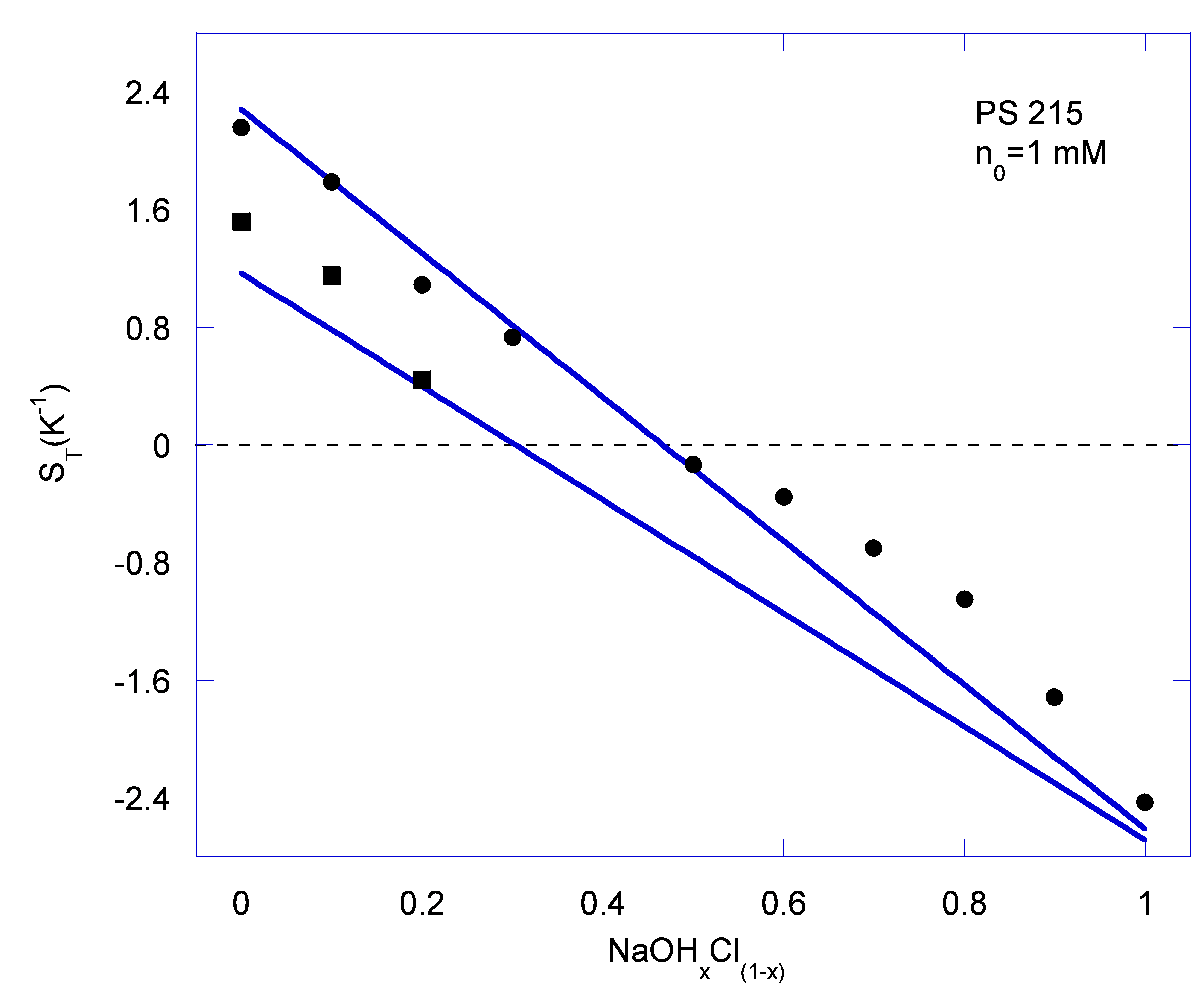}
\caption{Soret coefficient of 215 nm PS beads in a mixed electrolyte 
NaCl$_{1-x}$OH$_{x}$. The data at $41^\circ$
C (circles) show a linear dependence on $x$, and a change of sign at 
$x\approx \frac{1}{2}$; those at  $26^\circ$ C (squares) indicate that 
in NaCl solution, $S_{T}$ strongly increases
with temperature. The solid lines are calculated from Eq. (\protect\ref{6})
with $\protect\zeta =-38$ mV, $\protect\varepsilon =78\protect\varepsilon %
_{0}$, and $\protect\tau =1.4$; the fit values for the Seebeck coefficient $%
S $ and thermal diffusion factor $\protect\alpha $ compare favorably with
those measured at room temperature by Agar \protect\cite{Aga89}. As
discussed below, the temperature dependence of $S$ and $\protect\alpha $\
agrees with Eq. (\protect\ref{12}). }
\label{fig1}
\end{figure}

\begin{table}[b]
\caption{Seebeck coefficient $S$ and thermal diffusion factor $\protect%
\alpha $ of dilute solutions of NaCl and NaOH. The experimental values are
determined from the heats of transport $Q_{\text{Na}}^{\ast }=$3.5 kJ/mole, $%
Q_{\text{Cl}}^{\ast }=$0.5 kJ/mole, $Q_{\text{OH}}^{\ast }=$17.2 kJ/mole,\
measured by Agar \protect\cite{Aga89}, according to $\protect\alpha _{\text{%
NaA}}=(Q_{\text{Na}}^{\ast }+Q_{\text{A}}^{\ast })/2k_{B}T$ and $S_{\text{NaA%
}}=(k_{B}/e)(Q_{\text{Na}}^{\ast }-Q_{\text{A}}^{\ast })/2k_{B}T$. The fit
values are used for the theory curves in Fig.\ 2. These values agree with
the linear temperature dependence for NaCl solutions given in Eq. (\protect
\ref{12}). }%
\begin{tabular}{|c|c|c|c|c|}
\hline
& $\alpha _{\text{NaCl}}$ & $\alpha _{\text{NaOH}}$ & $%
\begin{array}{c}
S_{\text{NaCl}} \\ 
\text{(}\mu \text{V/K)}%
\end{array}%
$ & $%
\begin{array}{c}
S_{\text{NaOH}} \\ 
\text{(}\mu \text{V/K)}%
\end{array}%
$ \\ \hline
exp 25$%
{{}^\circ}%
\ $C & $0.8$ & $4.1$ & $52$ & $-238$ \\ \hline
fit 26$%
{{}^\circ}%
\ $C & $1.2$ & $4.8$ & $56$ & $-296$ \\ \hline
fit 41$%
{{}^\circ}%
\ $C & $2.9$ & $4.8$ & $138$ & $-296$ \\ \hline
\end{tabular}%
\end{table}

\section{Electrolyte composition}

\subsection{Specific-ion effects}

Fig. \ref{F2} shows the Soret coefficient of 215 nm PS\ beads in a 1 mM
solution of NaCl$_{1-x}$OH$_{x}$ as a function of the composition parameter $%
x$. The data show a linear variation with $x$; those at 41$%
{{}^\circ}%
$ C change of sign at $x\approx \frac{1}{2}$. These findings are in
agreement with previous studies reporting $S_{T}>0$ in NaCl and $S_{T}<0$ in
NaOH \cite{Put05,Vig10}; a similar linear dependence on composition was
reported for SDS micelles in a mixed electrolyte \cite{Vig10}. Moreover, the
data at 26$%
{{}^\circ}%
$ and 41$%
{{}^\circ}%
$ \ C indicate that the Soret coefficient in NaCl solution increases with
temperature.

The only ion-specific quantities in Eq. (\ref{6}) are the Seebeck
coefficient $S$ and the thermal diffusion factor $\alpha $; for a mixed
electrolyte we write these quantities as mole-fraction weighted averages,%
\begin{equation}
\begin{array}{c}
S=(1-x)S_{\text{NaCl}}+xS_{\text{NaOH}}, \\ 
\alpha =(1-x)\alpha _{\text{NaCl}}+x\alpha _{\text{NaOH}}.%
\end{array}
\label{8}
\end{equation}%
from the numbers given in Table I\ it is clear that $\alpha $ increases with 
$x$, whereas the Seebeck coefficient strongly decreases and changes sign.\
Moreover, with typical values of the surface potential of less than mV, one
finds that the composition dependence of Eq. (\ref{6}) is dominated by the
Seebeck term. Since most colloids carry a negative surface charge, one
expects quite generally a negative Soret coefficient in NaOH\ solution.

The fit curves are calculated with constant $\zeta $-potential and the salt
parameters given in Table I. We have neglected various additional
dependencies, such as the well known variation of the $\zeta $-potential
with the pH\ value. Indeed, our measurements (Stabino) show that, depeding
on salinity, $\zeta $ varies by about 10 to 30 percent upon replacing NaCl
with NaOH. Including this dependence in the fits would merely modify the
parameters $S$ and $\alpha $, but not affect our conclusions.

\subsection{Salinity}

In Fig. 3 we plot the Soret coefficient in NaCl solution as a function of
the salinity $n_{0}$. First $S_{T}$ increases with $n_{0}$, attains a
maximum at about 5 mM, and then drops to small values beyond the resolution
of our experiment. The\ behavior between 10 and 100\ mM arises from the
variation of the surface potential with salinity in the framework of Debye-H%
\"{u}ckel approximation, $\zeta \propto n_{0}^{-1/2}$. This law ceases to be
valid at lower salinity, as is well-known from electrophoresis \cite{Ant97}.

As a possible mechanisms for the dependence at low electrolyte strength, and
in particular for the occurrence of a maximum, we mention the surface
conductivity contribution, similar to well-known reduction of the
electrophoretic mobility \cite{Ant97}. On the other hand, there are several
effects specific for thermophoresis that have hardly been investigated so
far, such as the collective thermoelectric effect \cite{Maj11}, or the fact
that the coefficients $S$ and $\alpha $ depend significantly on salinity 
\cite{Cal73}.

In moderate or weak electrolytes, there is no satisfactory model for the
surface potential and its subtle dependencies on pH, temperature, and
salinity. In order to account for the variation of $S_{T}$ with salinity, we
take in all fits $\zeta =-38$ mV at 1\ mM, $\zeta =-53$ mV at 5\ mM, and $%
\zeta =-6.5$ mV at 100\ mM. The value at 5 mM agrees with the measured $%
\zeta $-potential over the temperature range studied here; those at 1 and
100\ mM are in accord with Fig. 3 and more generally with electrophoretic
mobility data.

\section{Temperature dependence\textit{\ }}

Soret data for various colloids show a strong and surprisingly universal
temperature dependence \cite{Iac03,Iac06,Put07,Nin08,Won12,Koe13}. Here we
present data for both NaCl and NaOH\ solutions that are essential for the
main results of this paper.

\begin{figure}
\includegraphics[width=\columnwidth]{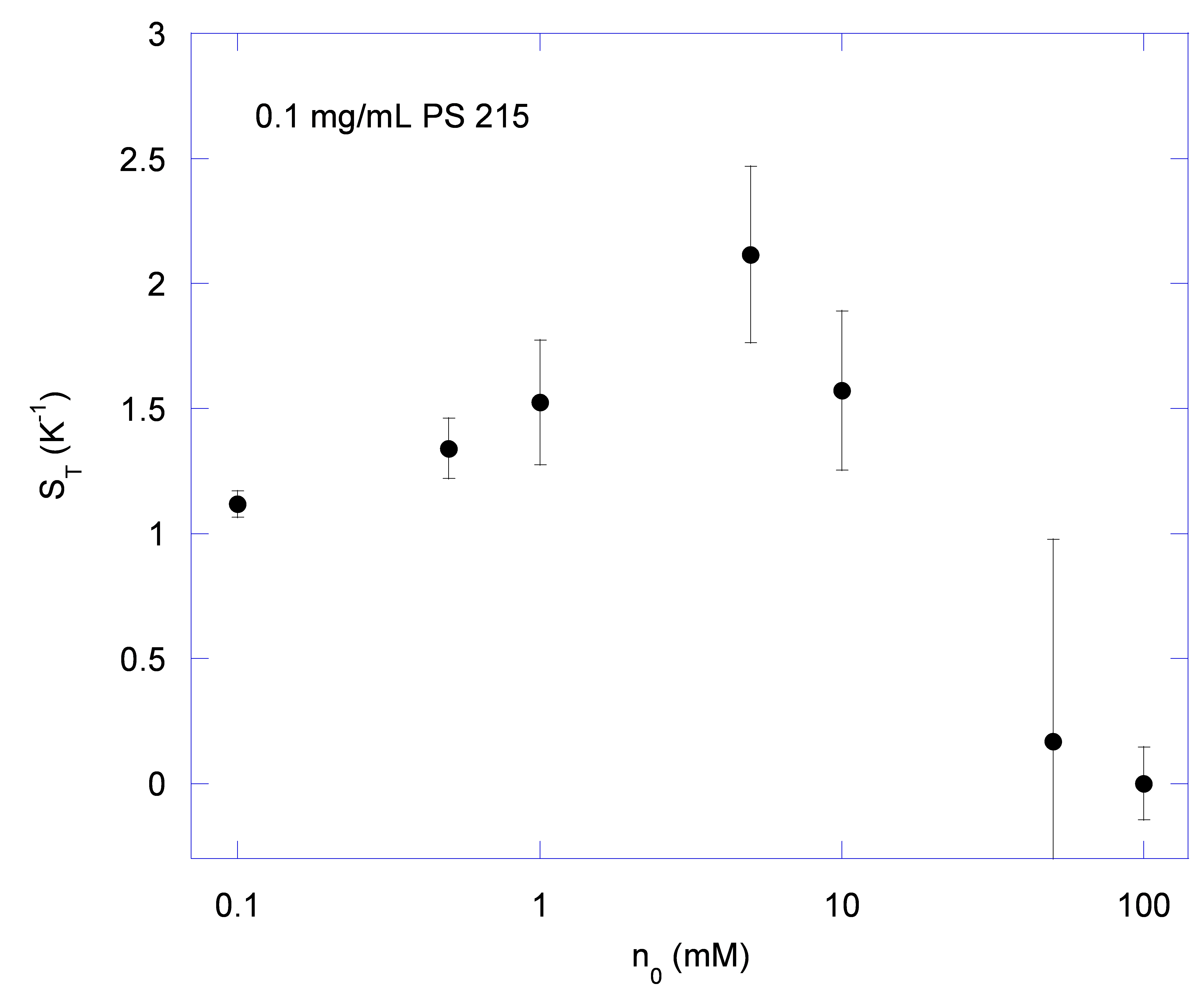}
\caption{Soret coefficient of 215 nm PS\
beads in NaCl solution as a function of the electrolyte strength at 
$T_{C}=299$K with $\Delta T=15$ K. For high salinity, $n_{0}>10$ mM,
inserting the Debye-H\"{u}ckel law $\protect\zeta \propto n_{0}^{-1/2}$ in 
(\protect\ref{6}) gives $S_{T}\propto n_{0}^{-1/2}$ and $\propto n_{0}^{-1}$
for the Seebeck term and the remainder, respectively; this agrees at least
qualitatively with the data. Yet these laws cease to be valid at lower
salinity; our data, and in particular the occurrence of a maximum, are
similar to the well-known behavior of the electrophoretic mobility, which is
to large extent determined by the salinity dependence of the 
$\protect\zeta$-potential \protect\cite{Ant97}.}
\end{figure}

\subsection{NaCl solution}

According to Fig.\ 3, the Soret coefficient of polystyrene particles
vanishes at high salinity. In order to determine whether this is the case at
all temperatures, we plot in Fig.\ 4 Soret data measured at different
electrolyte strength as a function of $T$. The main features are a roughly
linear variation with temperature, a strong reduction upon increasing the
electrolyte strength, and a linear increase with the particle size. These
findings lead us to two main conclusions of the present work.

\textit{First,} the Soret effect of 215 nm beads in $5$ mM\ NaCl is almost
ten times stronger than in $100$ mM solution, and this over the range from 25%
$%
{{}^\circ}%
$\ C\ to 50$%
{{}^\circ}%
$ C.\ This uniform reduction provides strong evidence that thermophoresis of
polystryene beads arises from charge effects only.\ They exclude the
existence of a significant non-ionic contribution to $S_{T}$; the values in $%
100$ mM solution may be taken as an upper bound. In contrast to our results,
the Soret effect of proteins arises essentially from a non-ionic mechanism,
as shown by a study on lysozyme at variable eletrolyte strength and
temperature \cite{Iac03}. This implies that thermophoresis of PS beads
differs fundamentally from that of proteins.

\begin{figure}
\includegraphics[width=\columnwidth]{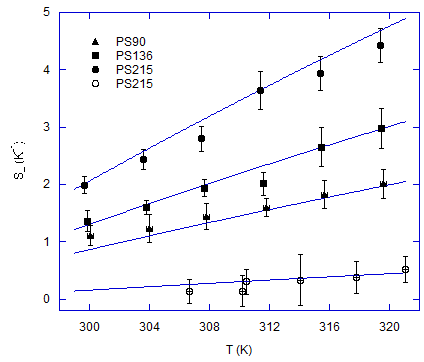}
\caption{Soret coefficient in NaCl solution as a function of
temperature. The solid lines are calculated from (\protect\ref{6}) with 
$\protect\zeta =-53$ mV at 5 mM and $-6.5$ mV at 100 mM; we assume a linear
variation of $S$ and $\protect\alpha $\ according to Eq. (\protect\ref{12}),
with values at 25 C.}
\end{figure}

\textit{Second,} the Soret coefficient at 46$%
{{}^\circ}%
$ C is about twice as large as at 26$%
{{}^\circ}%
$ C. Since the overall prefactor in (\ref{6}), and the parameters $\zeta $, $%
\varepsilon $, and $\tau $, hardly vary with temperature, we conclude that
the $T$-dependence of $S_{T}$ is necessarily related to the Seebeck
coefficient $S$ and the thermal diffusion factor $\alpha $ of the salt
solution. Moreover, the data confirm the well-known fact that the Soret
coefficient is proportional to the particle size \cite{Wue10,Bra08}. The
temperature dependence qualtitatively agrees with previous studies on PS
beads \cite{Iac06,Put07}.

In order to make the above arguments more quantitative, we have fitted the
temperature dependence in terms of a minimal model for the Seebeck
coefficient and the thermal diffusion factor, 
\begin{equation}
S_{\text{NaCl}}=S_{\text{NaCl}}^{0}\left( 1+\xi t\right) ,\ \ \ \alpha _{%
\text{NaCl}}=\alpha _{\text{NaCl}}^{0}\left( 1+\xi t\right) ,  \label{12}
\end{equation}%
where $t=T-298$K and the superscript \textquotedblleft 0\textquotedblright\
indicates room-temperature values according to Table I. The straight lines
in Fig.\ 4 are calculated from Eq. (\ref{6}) with the slope parameter $\xi
=0.14$ K$^{-1}$. There seem to be no data on the $T$-dependence of the
Seebeck coefficient.\ The slope measured for the thermal diffusion
coefficient $\alpha $ of very strong electrolytes of about 1 M, is about
four times smaller \cite{Cal73}; at low salinity a stronger but non-linear
increase of $\alpha $ was observed \cite{Gae82}. These experimental findings
are compatible with the model (\ref{12}) and the value for $\xi $.

\begin{figure}
\includegraphics[width=\columnwidth]{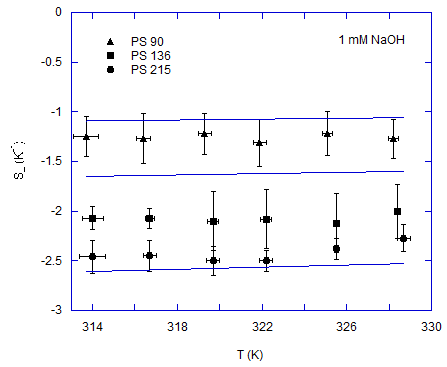}
\caption{ Soret coefficient of PS beads in NaOH
solution. The electrolyte thermal diffusion Seebeck coefficients $\protect%
\alpha $ and $S$\ are kept constant;\ the slight temperature dependence
arises from the prefactor in (\protect\ref{6}).}
\end{figure}

\subsection{NaOH solution}

In Fig.\ 5 we plot the Soret coefficient of PS beads of different size in
NaOH solution between 40$%
{{}^\circ}%
$ and 55$%
{{}^\circ}%
$ C. In this range $S_{T}$ varies little with temperature. The curves are
calculated from Eq. (\ref{6}) with $\zeta =-38$ mV and constant values for $%
S_{\text{NaOH}}$ and $\alpha _{\text{NaOH}}$, as given in Table I. The data
for beads of 90, 136, and 215 nm diameter follow roughly the linear size
dependence of Eq. (\ref{6}); the slight discrepancy might be related to the
surfactant-free surface, and thus slightly higher $\zeta $-potential, of the
homemade 90 and 136 nm sample, in contrast with the commercial 215 nm
beads.\ 

As a corollary, we note that our discussion of the data of Figs.\ 4 and 5
implies that solutions of NaCl and NaOH strongly differ in their thermal
diffusion behavior. The strong temperature dependence of Fig.\ 4 is related
to the variation of the coefficients in Eq. (\ref{12}), which are
qualitatively confirmed by experiment \cite{Cal73,Gae82}. The data of Fig.\
5 suggest that $S_{\text{NaOH}}$ and $\alpha _{\text{NaOH}}$ do not depend
on $T$; we are not aware of an experimental study on this point.

The coefficients $S$ and $\alpha $ result from the heat of transport $Q_{\pm
}$, which in turn are given by the ionic solvation enthalpy \cite%
{Aga89,Wue13}. Besides the electrostatic self-energy, these latter
quantities comprise van der Waals and hydration contributions which are not
easily evaluated. In the view of the well-known specific ion effects, as
expressed for example in the Hofmeister series for protein solutions \cite%
{Zha06}, it would not come as a surprise that NaCl and NaOH strongly differ
in their solvation enthalpies.

\section{Discussion}

Together with Eq. (\ref{6}) the data presented in the above figures show
that the Soret effect of PS beads originates essentially from thermal
diffusion of the salt and the electrolyte Seebeck effect, thus confirming
previous works \cite{Put05,Vig10}. As a most important novel finding, our
data show moreover that the Seebeck effect and diffusiophoresis in the salt
gradient lead to the strong temperature dependence in NaCl solution.
Although the variation with $T$ is similar to the behavior observed for
various ionic and non-ionic colloidal suspensions \cite{Iac06}, our
measurements give clear evidence that there are two distinct mechanisms at
work: Whereas a study on lysozyme protein shows that the temperature
dependence of the Soret data results from a non-ionic interaction \cite%
{Iac03}, the present work indicates that charge effects are dominant for
polystyrene beads.

So far all experimental studies reported an increase with temperature, which
is generally well fitted by the Piazza's phenomenological law $%
S_{T}=S_{T}^{\infty }(1-e^{(T^{\ast }-T)/T_{0}})$ that changes sign at $%
T=T^{\ast }$; extrapolating our NaCl data gives $T^{\ast }\approx 15%
{{}^\circ}%
$ C which roughly agrees with the values of \cite{Iac06,Put07}. On the other
hand, the data in NaOH solution presented in Fig. 5 provide the first
example for a $T$-independent Soret coefficient. This finding enforces our
above statement that the Soret motion of PS beads is a charge effect and
thus has a strikingly different temperature dependence.

There remains the question why the non-ionic mechanism at work in protein
and micellar solutions and the charge mechanism of latex particles in NaCl
solution, show such a similar temperature dependence. Piazza and coworkers
pointed out that the non-ionic interaction is strongly correlated with the
thermal expansivity $\beta $ of water \cite{Iac06}. On the other hand, a
similar correlation is known to occur between the thermal diffusion factor $%
\alpha _{\text{NaCl}}$ of sodium chloride and $\beta $ \cite{Cal73}. The
data of Fig.\ 4 suggest that there is a similar relation between the
electrolyte Seebeck coefficient $S_{\text{NaCl}}$ and the expansivity of
water. In this picture, the thermal expansion of water would be at the
origin of the temperature dependence of Soret data for non-ionic polymer
solutions, proteins, and charged PS particles. By contrast, the data of Fig.
5 indicate a very different behavior of NaOH\ solution.\ The physical origin
of such ion-specific effects \cite{Esl12,Lan12} could be related to the
Hofmeister effect observed for protein solutions.

\textbf{Acknowledgment.} A.W. acknowledges support through the Leibniz
program of Universit\"{a}t Leipzig during the summer term 2013, and thanks
the groups of Frank Cichos and Klaus Kroy for their kind hospitality.

\end{document}